\documentclass[twocolumn,amsmath,amssymb,superscriptaddress]{revtex4-1}

\usepackage{graphicx}
\usepackage{dcolumn}
\usepackage{bm}
\def\Vec#1{\mbox{\boldmath $#1$}}

\def\rv{\mbox{\boldmath $r$}}

\begin{document}


\title{Quasiparticle Bound States around Fractional Vortices in $s$-wave Superconductor }

\author{Yuki Nagai}
\affiliation{CCSE, Japan  Atomic Energy Agency, 178-4-4, Wakashiba, Kashiwa, Chiba, 277-0871, Japan}
\affiliation{
Mathematical Science Team, RIKEN Center for Advanced Intelligence Project (AIP), 1-4-1 Nihonbashi, Chuo-ku, Tokyo 103-0027, Japan
}

\author{Yusuke Kato}%
\affiliation{%
Department of Basic Science, University of Tokyo, Tokyo 153-8902, Japan}%

\date{\today}

\begin{abstract}
Since the experimental formation of a vortex with a fractional quantized magnetic flux in a thin superconducting $s$-wave bi-layer was recently reported, we calculate the local density of states (LDOS) around two half quantum vortices in a $s$-wave superconductor 
on the basis of the quasiclassical Eilenberger theory to investigate the effect of the phase discontinuity. 
We show that the LDOS pattern around fractional vortices is qualitatively different from that around conventional vortex. 
We find that the novel bow-tie-shaped bound states appear in the high energy region, which can be an evidence for the fractional vortices through scanning tunneling microscopy/spectroscopy. 
 We also show that the system with $n$ fractional vortices with the vorticity $1/n$ has a similarity to the Josephson junctions.
\end{abstract}

\maketitle
\section{Introduction}
Recently, Tanaka {\it et al.} reported observation of fractionalization of magnetic flux $\Phi_0$ in a thin superconducting Nb bi-layer \cite{Tanaka} through magnetic field distribution image taken by a scanning superconducting quantum interference device (SQUID) microscope. As an earlier relevant study, Moler's group reported signature for phase soliton \cite{TanakaPRL} and fractionally quantized flux in the pattern of current as a functional of the applied flux in superconducting aluminum rings of various sizes \cite{Moler2006}. Her group also reported sub-$\Phi_0$ features in highly underdoped cuprate superconductors YBCO and they attributed it to the kinked stacks of pancake vortices\cite{Moler2008}.
The authors in Ref.~\cite{Tanaka} claimed that fractional vortices in their system are explained by a multi-component $s$-wave conventional superconductivity. 

The term ^^ ^^ fractional vortex", however, has been used in two ways: fractionally-quantized-flux vortex and/or fractional-phase vortex.
The former is the vortex whose magnetic flux is fractionally quantized and 
the latter is the vortex whose superconducting phase is fractional.  
In both cases, the magnetic flux of a vortex is fractionally quantized. 
Fractionally-quantized-flux vortices have been discussed in multi-component superconductors\cite{TanakaPRL,Babaev,Lin}. 
Machida {\it et al.} suggested that half fractionally-quantized-flux vortices spontaneously emerge in corners of $d$-wave superconductor 
dot embedded in $s$-wave superconducting matrix due to the phase interference between superconducting order parameters with different symmetries.\cite{Machida} 
Fractional-phase vortices has been discussed by Volovik in chiral superfluids and superconductors\cite{Volovik}. 
The half quantum phase vortices are allowed in the triplet superfluids due to their spin degrees of freedom. 
The stability of the half quantum vortices have been discussed in $p_{x} + i p_{y}$ superconductors by Chung {\it et al.}\cite{Chung} 
They showed that the stability of two half quantum vortices is enhanced when the ratio of spin superfluid density to superfluid density is small.

The observation of fractional quantized magnetic flux in the bi-layer $s$-wave superconductor does not necessarily mean the existence of 
the latter fractional vortex, a fractional-phase one, since the SQUID experiment can not measure the phase of the order parameter. 
Thus, another experimental measurement is needed to conclude which fractional vortex appears in the bi-layer $s$-wave superconductor. 
Through out this paper, we use ^^ ^^ fractional vortex" as the fractional-phase vortex.  

The quasiparticle bound states around a vortex reflects the pairing symmetry.\cite{Caroli,Ichioka,Hayashi,Nagai,Nagaitopo,Bao} 
The local density of states (LDOS) around a vortex has been investigated in various superconductors such as $s$-wave, 
anisotropic $s$-wave and $d$-wave superconductors. 
The LDOS can be observed by the scanning tunneling microscopy/spectroscopy (STM/STS) \cite{Hess,Nishimori,Tao}.
In the system with conventional vortices, the phase of the pair-potential changes $2 \pi$ around a vortex, 
so that quasiparticles feel $\pi$-shift of pair-potential running through a vortex core. 
In the system with the half quantum vortices, 
there is the phase discontinuity, so that 
the shape of the quasiparticle bound states around these fractional vortices is nontrivial.

\begin{figure}
  \begin{center}
    \begin{tabular}{p{40mm}p{40mm}}
      \resizebox{38mm}{!}{\includegraphics{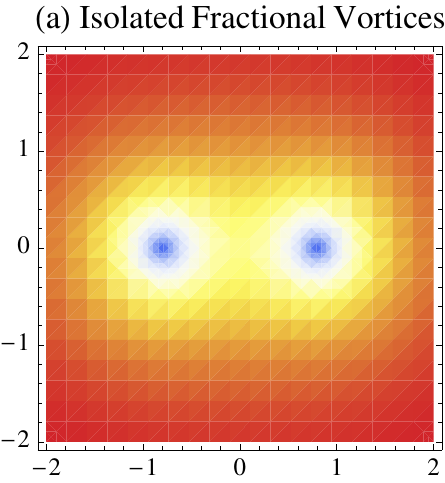}} &
      \resizebox{38mm}{!}{\includegraphics{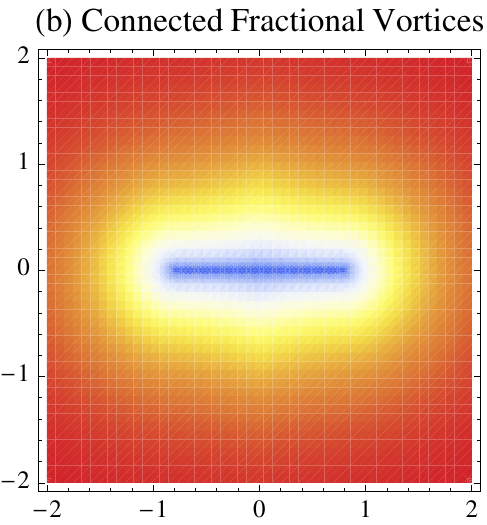}} 
    \end{tabular}
\caption{\label{fig:gap}
(Color online)
Schematic figures of a spatial dependence of the gap amplitude $f(\Vec{r})$ with two half vortices. (a)``Isolated'' denotes that 
$f(\Vec{r}) \neq 0$ on the branch cut. (b) ``connected'' denotes that $f(\Vec{r}) = 0$ on the branch cut.
}
  \end{center}
\end{figure}

The purpose of this paper is to investigate the quasiparticle bound states around fractional vortices in $s$-wave superconductor. 
For simplicity, 
we consider the single-band isotropic $s$-wave superconductor with the two-dimensional isotropic Fermi surface on the basis of 
quasiclassical Eilenberger theory\cite{Eilenberger,Larkin}. 
It is appropriate to investigate the phase discontinuity. 
We consider two kinds of the system with half vortices. 
The former is the system with ^^ ^^ isolated'' fractional vortices, where the gap amplitude on the line between 
two half vortices is not zero (See, Fig.~\ref{fig:gap}(a)). 
The latter is the system with ``connected'' fractional vortices, where the gap amplitude on the line between 
two half vortices is zero (See, Fig.~\ref{fig:gap}(b)). 
We discuss how to obtain the evidence for the fractional vortices. 

This paper is organized as follows. 
The model and method are shown in Sec.~II. 
The systems with the isolated fractional vortices are shown in Sec.~III. 
In this section, we consider the two half vortices and $n$ fractional vortices with vorticity $1/n$. 
The system with the two connected fractional vortices  is shown in Sec.~VI. 
The discussion is given in Sec.~V. 
The conclusion is given in Sec.~VI.

\section{Model and Method}

We use the quasiclassical theory developed by Eilenberger, Larkin and
Ovchinnikov \cite{Eilenberger,Larkin}, which is suited to discuss most superconductors. We consider two-dimensional clean superconductors with isotropic Fermi surface in the type II limit for simplicity and we ignore the self-energy part of Green function and the vector potential.

We discuss LDOS $\nu(\epsilon,{\bm{r}})$ for quasiparticles with excitation energy $\epsilon$ at each position $\bm{r}$ around fractional vortices. For our purpose, it suffices to calculate diagonal part (in particle-hole space) of quasiclassical Green function $g(\epsilon+{\rm i}0,\hat{\bm{k}})$ as a function of $\epsilon$ and the position on the Fermi surface (denoted by a two-dimensional unit vector $\hat{\bm{k}}$). The LDOS follows from $g$ averaged over the Fermi surface $\langle \cdots\rangle_{\hat{\bm{k}}}$,
\begin{equation}
\nu(\rv,\epsilon) = - \nu(0) \langle{\rm Re} \; g(\epsilon+{\rm i}\delta,\hat{\bm{k}}) \rangle_{\hat{\bm{k}}}, \label{eq:ldos}
\end{equation}
where $\nu(0)$ denotes the density of states on Fermi surface 
in the normal metallic state.

The equation of motion for $g$ is called Eilenberger equation, which can be simplified by introducing auxiliary functions $a$ and $b$ defined by 
\begin{equation}
g = - \frac{1-a b}{1+ a b}. \label{eq:greengf}
\end{equation}
The auxiliary functions $a$ and $b$ satisfy 
\begin{eqnarray}
\Vec{v}_{\rm F} \cdot \Vec{\nabla} a + 2 \omega_n  a + a \Delta^{\ast}(\bm{r}) a - \Delta(\bm{r}) &=& 0, \label{eq:ar}\\
\Vec{v}_{\rm F} \cdot \Vec{\nabla} b + 2 \omega_n  b - b \Delta(\bm{r}) b + \Delta^{\ast}(\bm{r}) &=& 0, \label{eq:br}
\end{eqnarray}
which is Eilenberger equation in Riccati formalism\cite{Nagato,Schopohl}.
Here $\Vec{v}_{\rm F}=v_{\rm F}\hat{k}$ denotes the Fermi velocity and $\Delta(\bm{r})$ the gap function.
We use the analytic continuation $i \omega_n \rightarrow \epsilon+{\rm i}\delta$ when we calculate the LDOS. 
Since these equations (\ref{eq:ar}) and (\ref{eq:br}) contain $\Vec{\nabla}$ only through $\Vec{v}_{\rm F} \cdot \Vec{\nabla}$, 
they reduce to a one-dimensional problem on a straight line, 
the direction of which is given by that of the Fermi velocity $\Vec{v}_{\rm F}$.
We calculate the LDOS by solving (\ref{eq:ar}) and (\ref{eq:br}) with the fourth order Runge-Kutta method.
For simplicity, we defined the gap function $\Delta(\Vec{r})$ without self-consistent gap equation.

There are branch cuts between fractional vortices. 
Crossing the branch cut, the phase $\arg( \Delta(\Vec{r})) \equiv \phi$ changes $\pi/n$. 
Here, $n$ denotes an inverse of a vorticity (i.e., $n = 2$ with two 
half vortices).  
We consider two cases about a spatial dependence of the gap amplitude. 
In the former case, we consider ``isolated'' fractional vortices, 
where the gap amplitude $|\Delta(\Vec{r})| \equiv f(\Vec{r}) \neq 0$ 
on the branch cut between each isolated fractional vortex 
as shown in Fig.~(\ref{fig:gap})(a). 
In the latter, we consider ``connected'' fractional vortices, 
where 
$f(\Vec{r}) = 0$ on the branch cut between each connected fractional vortices 
as shown in Fig.~(\ref{fig:gap})(b).

\section{Isolated Fractional Vortices}
\subsection{Two Half Vortices}
\begin{figure}
\includegraphics[width = 5cm]{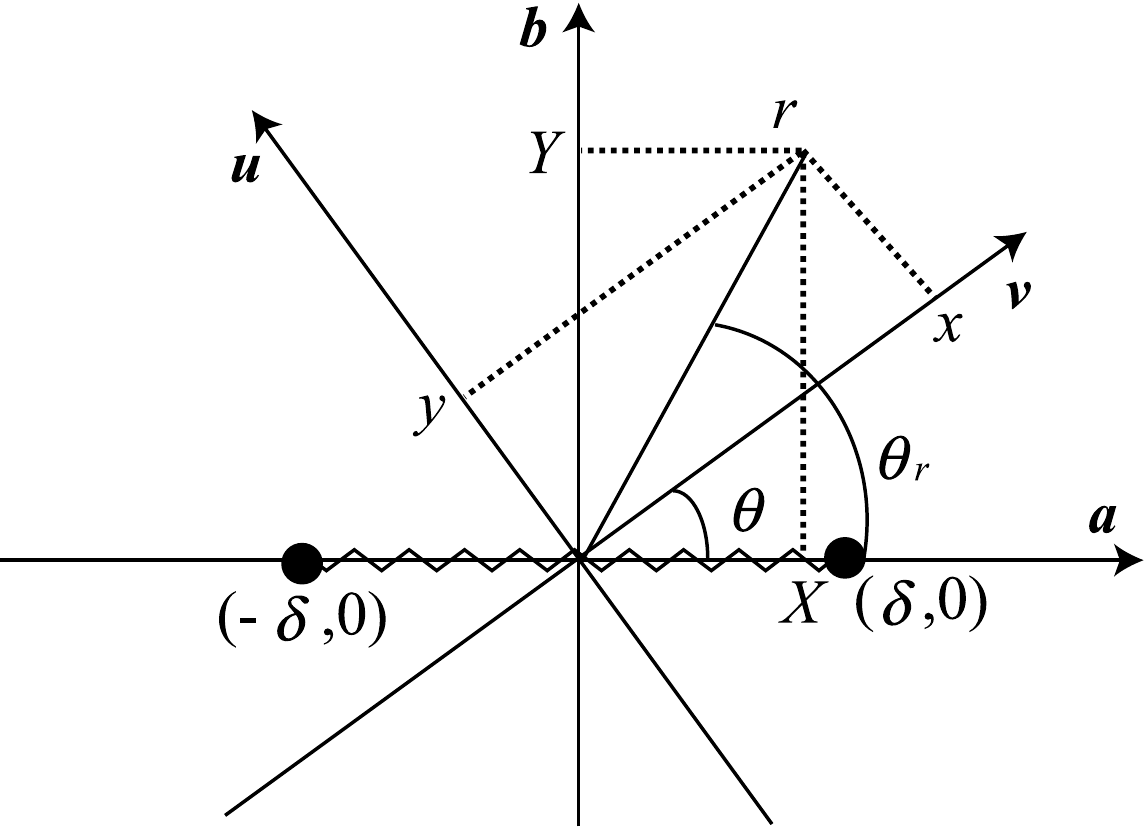}
\caption{\label{fig:fzahyo} Frame of the space with two half vortices. $\delta$ is the inter-vortex distance. 
$\theta_r$ is the angle between the $a$ axis and $\Vec{r}$.}
\end{figure}
We consider the system with the two isolated half fractional vortices. 
Here, the origin is at a midpoint between half vortices(See, Fig.~\ref{fig:fzahyo} and our previous paper\cite{Nagai}). 
We denote by $X, Y$ a Cartesian coordinate in a laboratory frame (and we reserve the symbols $x$ and $y$ as the coordinate for $\hat{\bm{k}}$ dependent-frame). 
We consider the pair function of the two isolated half fractional vortices written as 
\begin{eqnarray}
\frac{\Delta(r)}{\Delta_{\infty}} &=& f_+(X,Y) f_-(X,Y) \exp[i (\phi_+(X,Y)+\phi_-(X,Y))], \nonumber \\ \\
	&\equiv&  f(r) \exp [i \phi(r)], \label{eq:gapf}
\end{eqnarray}
where 
\begin{eqnarray}
f_{\pm}(X,Y) &=&\tanh^{1/2} (\sqrt{(X \pm \delta)^2 + Y^2}), \\
\phi_{\pm}(X,Y) &=& \arctan(Y/(X \pm \delta))/2, 
\end{eqnarray}
as shown in Fig.~\ref{fig:gap}(a). $\Delta_\infty$ denotes the modulus of the gap function far away from the vortices. Here, $\delta$ denotes the inter-vortex distance. 
The phase $\phi(r)$ changes $\pi/2$ crossing the branch cut between two half vortices. 

\begin{figure}
\includegraphics[width = 7cm]{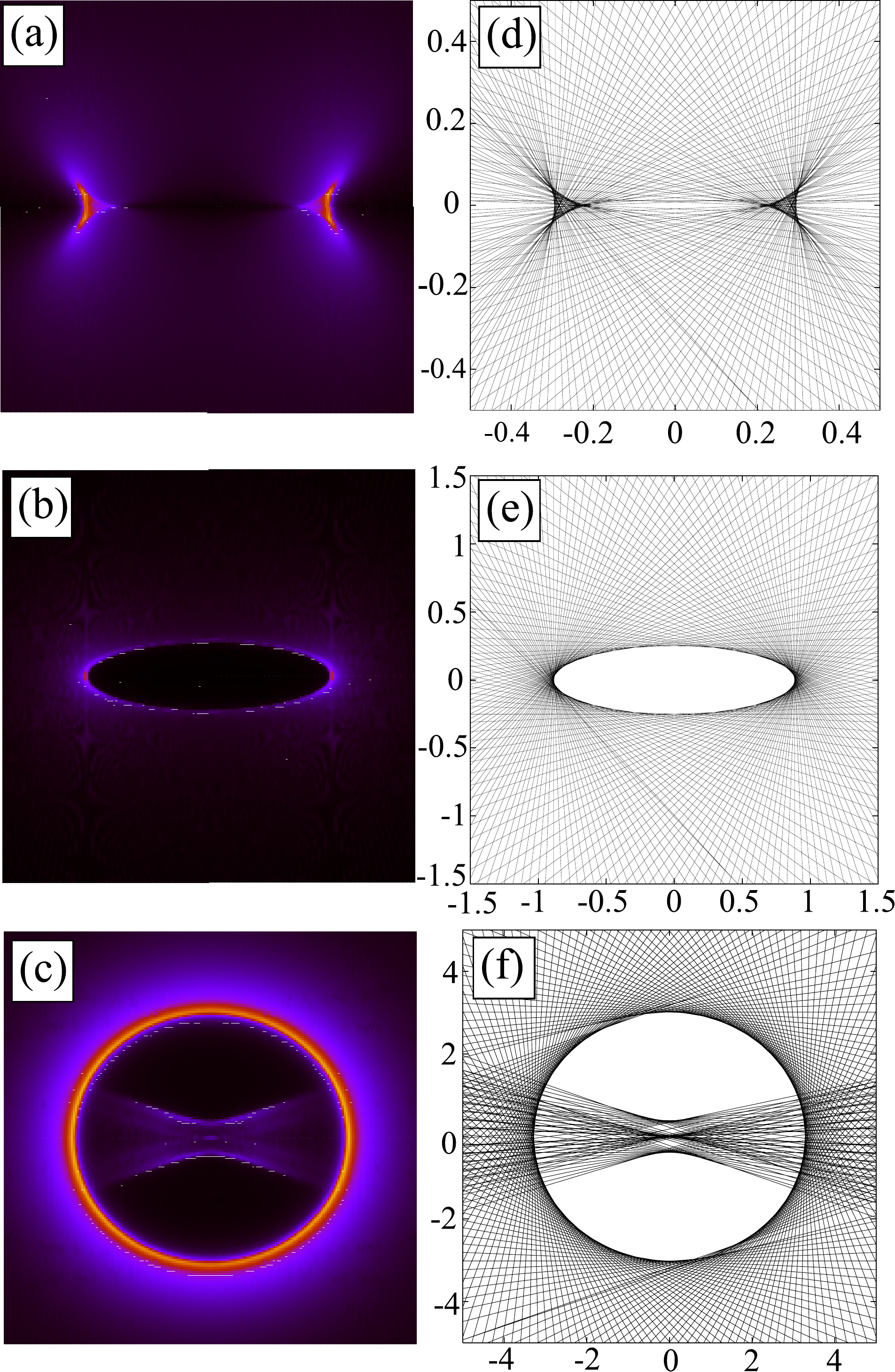}
\caption{\label{fig:fig2}(Color Online) Local density of the states in the system with two isolated half 
fractional vortices. 
The inter-vortex distance $\delta = 0.8 \xi_0$. 
The distributions of the LDOS on the left hand side and the sets of the quasiparticle paths on the right hand side. 
(a) and (d):the energy $\epsilon = 0.1 \Delta_{\infty}$, (b) and (e): the energy $\epsilon = 0.4 \Delta_{\infty}$
 and (c) and (f):the energy $\epsilon = 0.9 \Delta_{\infty}$. the smearing factor $\eta = 0.001\Delta_{\infty}$.}
\end{figure}
\subsubsection{Numerical Results}
We calculate the LDOS with $\delta = 0.8 \xi$ numerically as shown in Fig.~\ref{fig:fig2}(a)-(c). 
The LDOS patterns can be divided into three energy regions. 
In the low energy region as shown in Fig.~\ref{fig:fig2}(a) ($\epsilon \gtrsim 0.3 \Delta$), the LDOS pattern is quite different from that around a single vortex. 
The LDOS around a single vortex has a circular pattern and the peak position of this LDOS is located near 
the center of the vortex core. 
In contrast, the peak positions of the LDOS with the half vortices are located at $(X,Y) \sim (\pm 0.3 \xi_{0},0)$, 
whose points are near the midpoint between two half vortices. 
In the middle energy region ($0.3 \Delta \gtrsim \epsilon \gtrsim 0.8 \Delta$), 
the LDOS patterns change from that like the triangular to that like the ellipse. 
We show the LDOS with the energy $\epsilon = 0.4 \Delta$ in Fig.~\ref{fig:fig2}(b). 
In the high energy region, the novel quasiparticle bound states are found inside 
the conventional circular LDOS pattern (See, Fig.~\ref{fig:fig2}(c)).

In the quasiclassical theory, the equations of motion for quasiparticles running in a certain direction can be 
respectively solved since we assume the non-selfconsistent gap function as shown in Eqs.~(\ref{eq:ar}), (\ref{eq:br}) and (\ref{eq:gapf}). 
Therefore, results can be respectively explained by the enveloping curves of the set of quasiclassical paths.\cite{Nagai}
We draw the set of quasiclassical paths as shown in Fig.~\ref{fig:fig2}(d)-(f). 
We can show that the analysis by the quasiparticle paths used in our previous paper\cite{Nagai} can be applied 
to the analysis of the LDOS by the numerical method.
In the following sections, we explain the origins of the LDOS patterns in the each energy region.
\subsubsection{Low Energy Region}
\begin{figure}
\includegraphics[width = 8cm]{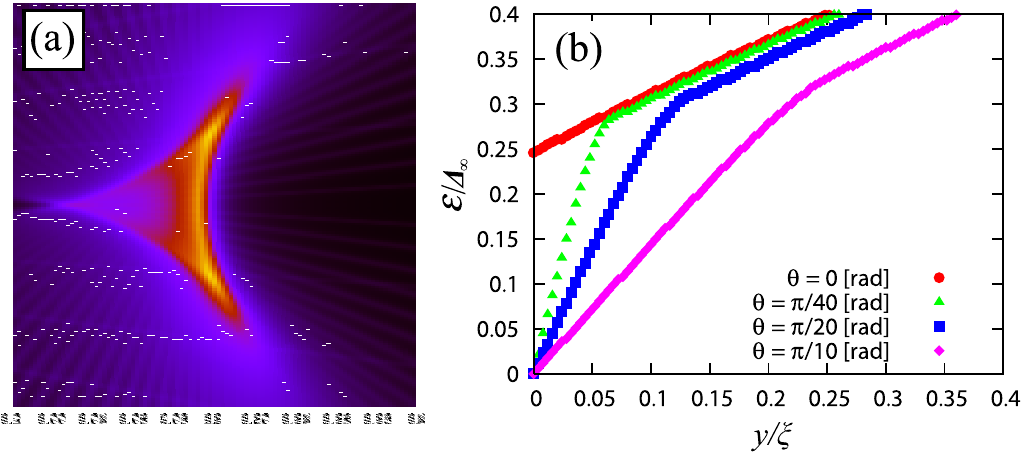}
\caption{\label{fig:ep01}(Color Online)(a): The local density of the states in the system with two isolated half 
fractional vortices at the energy $\epsilon = 0.1 \Delta_{\infty}$. This figure is the enlargement of Fig.~\ref{fig:fig2}(a). 
(b)The energy dependence of the peak positions for quasiparticles in the various directions.}
\end{figure}

The features of the LDOS pattern in this energy region are that 
the low energy LDOS pattern is like a triangular shape (See, Fig.~\ref{fig:ep01}(a)) and that 
the peak positions of the LDOS with the half vortices are located near the midpoint between two half vortices. 
Around a single vortex, the LDOS has a circular symmetric pattern and the peak position of this LDOS is located near 
the center of the vortex core. 

\begin{figure}[htbp]
 \begin{center}
  \includegraphics[width=6cm,keepaspectratio]{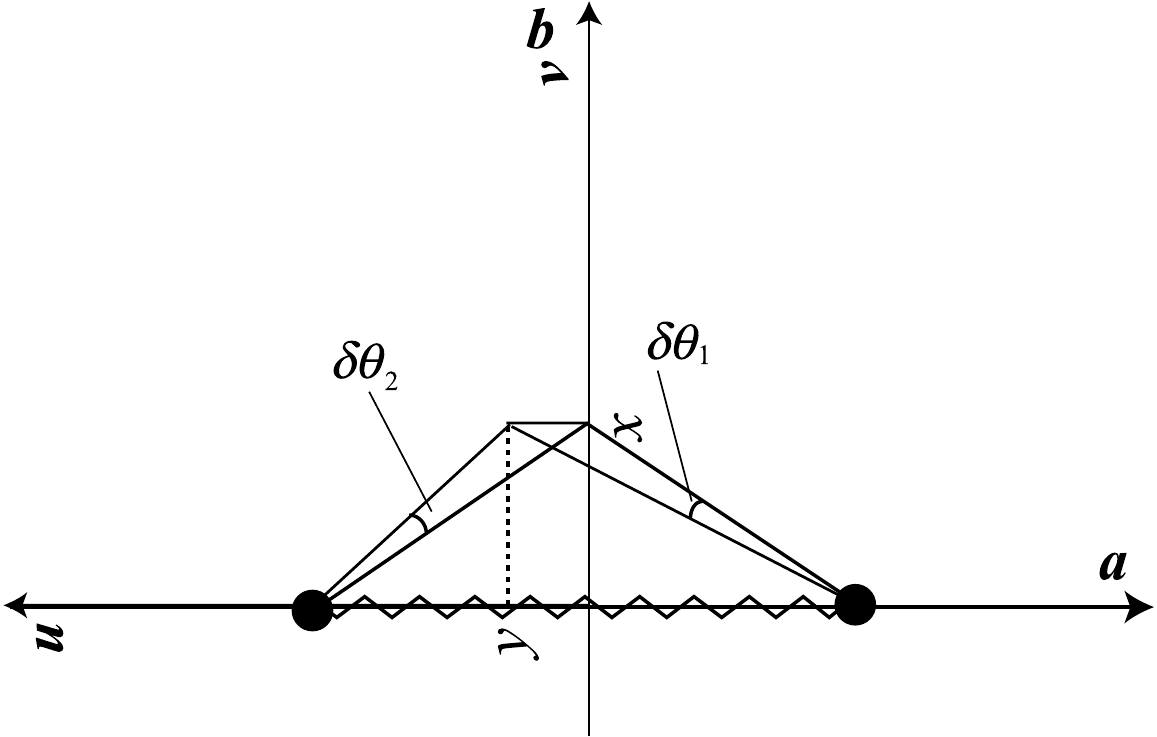}
\caption{Frame of the space with $\theta = \pi/2$.}
\label{fig:deltheta}
 \end{center}
\end{figure}

\begin{figure}
\includegraphics[width = 8cm]{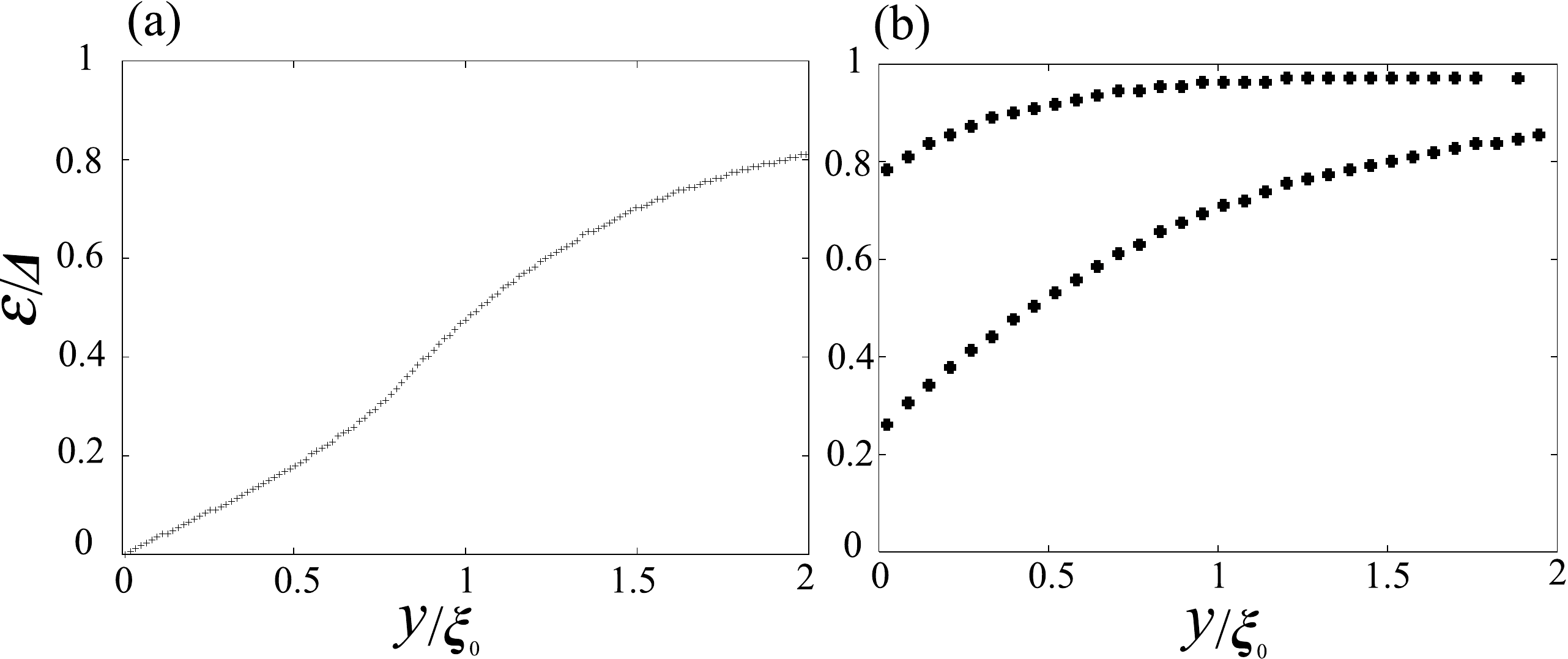}
\caption{\label{fig:fig3}Energy dependence of peak positions for the quasiparticle running in the (a):$Y$ direction and (b):$X$ direction with isolated two half vortices.}
\end{figure}

We show the energy dependence of the peak positions for the quasiparticles 
running in the $Y$-direction are shown in Fig.~\ref{fig:fig3}(a). 
$y$ in Fig.~\ref{fig:fig3}(a) denotes an impact parameter, which is equal to $X$ as shown in Fig.~\ref{fig:deltheta}. 
Therefore, these peak positions shift away from the origin with increasing the energy.

This feature is explained on the basis of the Kramer-Pesch approximation, 
which is appropriate in the low energy region.\cite{Nagai,Kramer} 
This approximation is the perturbation theory where the denominator of the Green function around the origin 
is expanded with respect to $y$ and an energy $\omega_n$.  
On the $Y$-axis, we can easily solve Eqs.~(\ref{eq:ar}),(\ref{eq:br}), since 
the phase of the pair-function $\phi(0,Y)$ changes only at the origin from $3 \pi/2$ to $\pi/2$. 
The Green function at the origin is written as 
\begin{eqnarray}
g(\Vec{r}=0; i \omega_n) &=& -\frac{1- a(x \rightarrow - \infty)b(x \rightarrow  \infty) }
{1 + a(x \rightarrow - \infty)b(x \rightarrow  \infty)}, \\
 &=&
  -\frac{(\omega_n + \sqrt{\omega_n^2 +|\Delta_{\infty}|^2})^2 + |\Delta_{\infty}|^2}
{(\omega_n+\sqrt{\omega_n^2 + |\Delta_{\infty}|^2})^2 - |\Delta_{\infty}|^2}. \: \: \: \: \: \: \:  
\end{eqnarray}
This Green function diverges at the zero energy ($i \omega_n = 0$) so that 
the Andreev bound states occur at the zero energy at the origin. 
We consider this Green function as an unperturbed Green function. 
Let us expand the pair function $\Delta(x,y) = f(x,y) \exp(i \phi(x,y))$ with respect to $y$ near the origin. 
We can regard as $f(x,y) \sim const.$ when the inter-vortex distance is large. 
The phase $\phi$ is expressed as 
\begin{eqnarray}
\phi(x > 0) &=& \frac{3 \pi}{2} + \frac{\delta \theta_{1}+\delta \theta_{2}}{2}, \\
\phi(x < 0) &=& \frac{ \pi}{2} - \frac{\delta \theta_{1}+\delta \theta_{2}}{2}. 
\end{eqnarray}
Here, $\delta \theta_{1}$ and $\delta \theta_{2}$ are defined as shown in Fig.~\ref{fig:deltheta}. 
In the first order with respect to $y$, $\exp (i \phi (x))$ is written as 
\begin{equation}
e^{i \phi (x)} \sim -{\rm sign}\:(x) \exp \left( {\rm i} \frac{ \pi}{2} \right) 
	\left(
        1 + {\rm i} \frac{x}{\delta^2 + x^2}y
         \right).
\end{equation}
Then, we obtain the pair function $\Delta(x,y)$ with respect to $y$: 
\begin{eqnarray}
\Delta(x,y) &\sim& - \Delta_{\infty} {\rm sign}\:(x) \exp \left( {\rm i} \frac{ \pi}{2} \right) 
	\left(
        1 + {\rm i} \frac{x}{\delta^2 + x^2}y
         \right),\nonumber \\ \\
        &=& \Delta_0 + {\rm i} \frac{x}{\delta^2 + x^2}y \Delta_0.
\end{eqnarray}
Here, $\Delta_{0} = - \Delta_{\infty}  {\rm sign}\:(x) \exp \left( {\rm i} \frac{ \pi}{2} \right)$. 
In the Kramer-Pesch approximation about a single vortex, the first order of the pair function is expressed as 
$\Delta_{1} = i y/x \Delta_{0}$. 
Therefore, as in the Ref.~\onlinecite{Nagai}, the energy dependence of the peak position for the quasiparticles running in the $Y$-direction 
is written as 
\begin{equation}
 y \propto \epsilon.
\end{equation}
This relation means that, with increasing the energy, the quasiparticle path shifts away from the origin, so that the peak positions of quasiparticle bound states also shift away from the origin on the $X$-axis.

As shown in Fig.~\ref{fig:ep01}(a), the low energy LDOS pattern is like a triangular shape. 
This shape is explained by the set of the quasiparticle paths shown in Fig.~\ref{fig:fig2}(d). 
The quasiparticles can not run in the $X$-direction, since the quasiparticle path in the $X$-direction 
does not exist as shown in Fig.~\ref{fig:fig2}(d). 
In the energy range $\epsilon < 0.3 \Delta_{\infty}$, the quasiparticles running such a direction is not allowed, 
since these quasiparticles have the gap as shown in Fig.\ref{fig:fig3}(b). 
We show the energy dependence of the peak positions for quasiparticles running in the various directions ($\theta$) 
in Fig.~\ref{fig:ep01}(b). 
In the energy $\epsilon = 0.1 \Delta_{\infty}$, the impact parameters $y$ for quasiparticles in Fig.~\ref{fig:ep01}(b) 
are small.  
Therefore, the quasiparticles can not form the circler symmetric LDOS pattern. 
The dispersion relation of the quasiparticle running in the $X$-direction shown in Fig.\ref{fig:fig3}(b).  
In Figs. \ref{fig:fig2}(d)-(f), the set of quasiparticle paths are drawn 
calculating the energy dispersion relations of the quasiparticles running in the various directions. 
\subsubsection{Middle Energy Region}
\begin{figure}[htbp]
 \begin{center}
  \includegraphics[width=6cm,keepaspectratio]{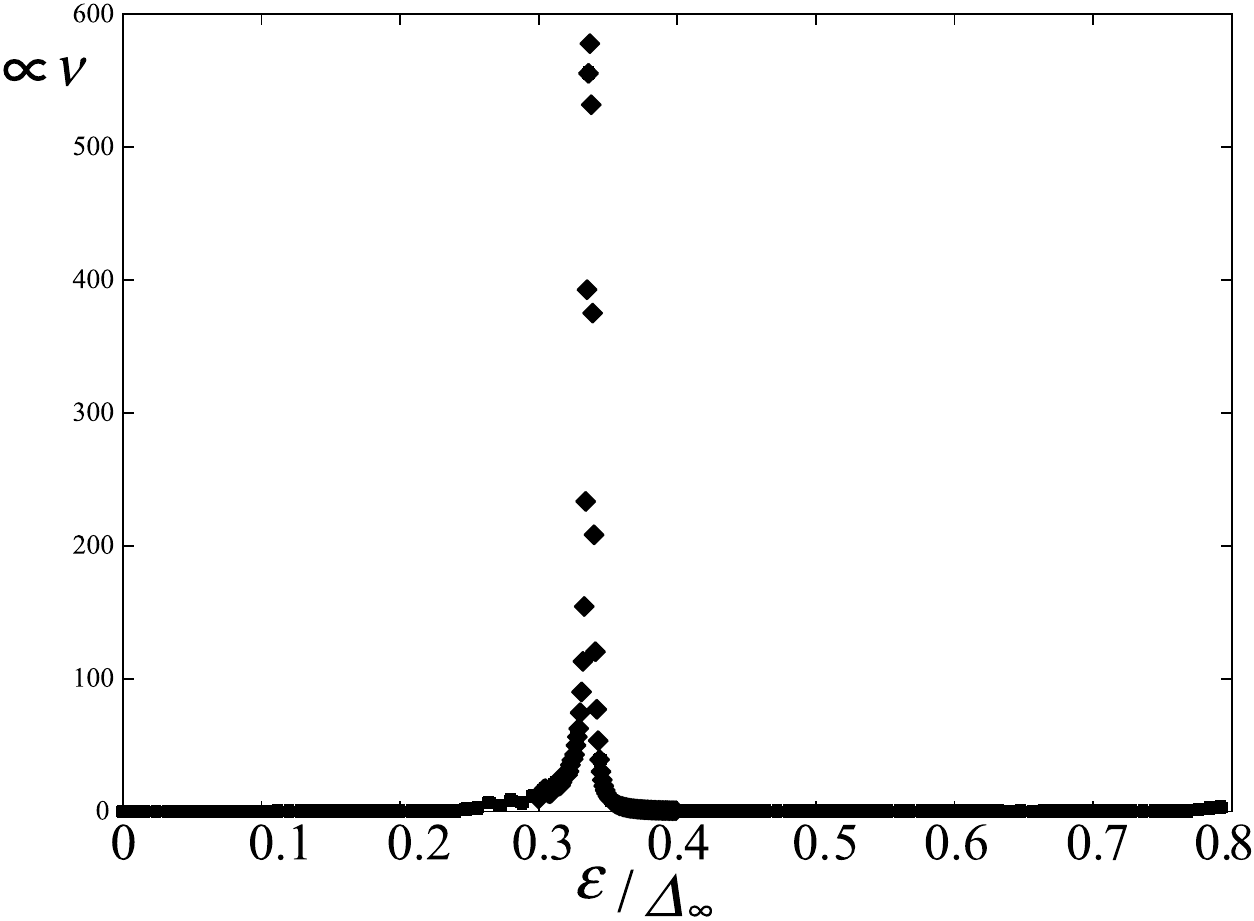}
\caption{Energy dependence of the LDOS at the half fractional vortex core.}
\label{fig:epdep}
 \end{center}
\end{figure}

At the gap energy $\epsilon \sim 0.3 \Delta$, the LDOS patterns change from that like the triangular to that like the ellipse. 
We show the LDOS with the energy $\epsilon = 0.4 \Delta$ in Fig.~\ref{fig:fig2}(b). 
In this middle energy region, the quasiparticles do not cross the branch cut ($- \delta < X < \delta$) as shown 
in Fig.~\ref{fig:fig2}(e). 
On the other hand, the quasiparticles cross the branch cut  as shown 
in Fig.~\ref{fig:fig2}(d) in the low energy region.  
We show the energy dependence of the LDOS at the half fractional vortex core ($X = \delta = 0.8 \xi$) in Fig.~\ref{fig:epdep}. 
In the energy $\epsilon = 0.33 \Delta$, the LDOS at the half vortex core diverges. 
This result shows that the peak position cross a vortex core with increasing energy. 

As shown Fig.~\ref{fig:ep01}(b), the energy dependence of peak positions in the various directions in 
this middle energy region 
can be written as 
\begin{equation}
\epsilon \sim c_{1} y + c_{0}.
\end{equation}
Here, $c_{0}$ and $c_{1}$ do not depend on the angle $\theta$. 
Therefore, the ellipsoidal LDOS patterns are in proportion to the energy in this energy region.
\subsubsection{High Energy Region}
In the high energy region, the novel quasiparticle bow-tie-shaped bound states are found inside 
the conventional circular symmetric LDOS pattern (See, Fig.~\ref{fig:fig2}(c)).
As shown in Fig.~\ref{fig:fig3}(b), another branch in the high energy region appears in 
the energy dependence of peak positions for the quasiparticles running in the $X$ direction. 
We show the energy dependence of the retarded Green function in the $X$ direction in Fig.~\ref{fig:fig4}(a). 
In this figure, two Andreev bound states are clearly observed under the gap energy. 
The bow-tie-shaped Andreev bound states with a high energy are near the origin as shown in Figs.~\ref{fig:fig2}(c) and (f).
The LDOS pattern in the high energy region are the superposition of the bow-tie-shaped pattern and the 
circular pattern, since the quasiparticle paths of the each pattern are superposed as shown in Figs.~\ref{fig:fig2}(f).

One can observe the LDOS pattern as the evidence of the fractional vortices in the high energy region rather than the low energy region.
In the low energy region, the novel bound states are located between two half vortices, so that one has to observe it precisely. 
One can easily observe the bow-tie-shaped LDOS patterns in high energy region by STM/STS 
since the novel bound states in this region are larger than the two half vortices.
\subsubsection{Dependence of the inter-vortex distance}

\begin{figure}
\includegraphics[width = 8cm]{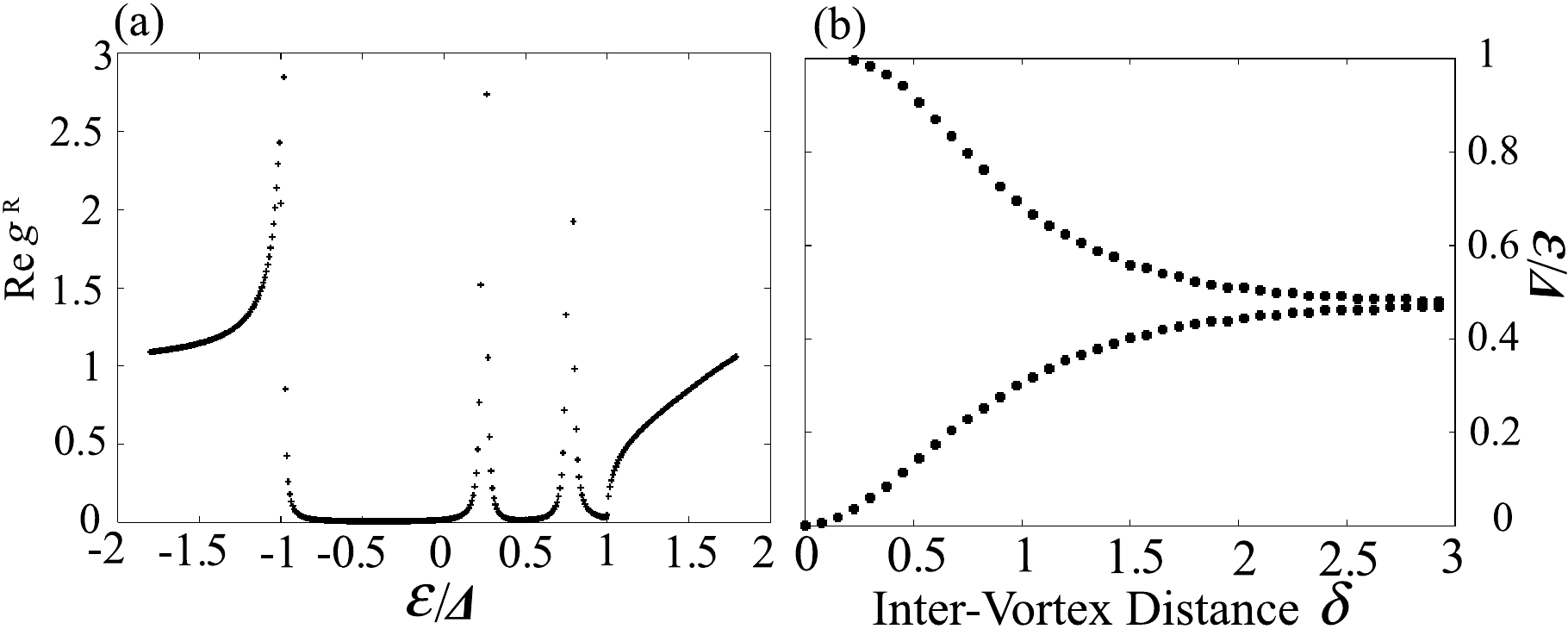}
\caption{\label{fig:fig4}(a):Energy dependence of the retarded Green function in the $X$-direction near the origin. 
(b) the inter-vortex distance dependence of peak positions for the quasiparticles in the $X$-direction. }
\end{figure}

The inter-vortex distance $\delta$ depends on the free-energy in the real materials. 
We investigate the energy dependence of peak positions for the quasiparticles running in $X$-direction as shown in Fig.~\ref{fig:fig4}(b) 
with increasing the inter-vortex distance. 
In the conventional single vortex limit ($\delta \rightarrow 0$), there is the one conventional Andreev bound states at the zero energy. 
With increasing the inter-vortex distance $\delta$, the energy of the novel bound states become small and the energy of the lower bound states become large. 
In the large inter-vortex distance limit ($\delta \rightarrow \infty$), 
these two dispersion curves merge. 
These results suggest that the high-energy novel bound states  appear in the region $\delta > 0.25 \xi$. 

\begin{figure}
\includegraphics[width = 5cm]{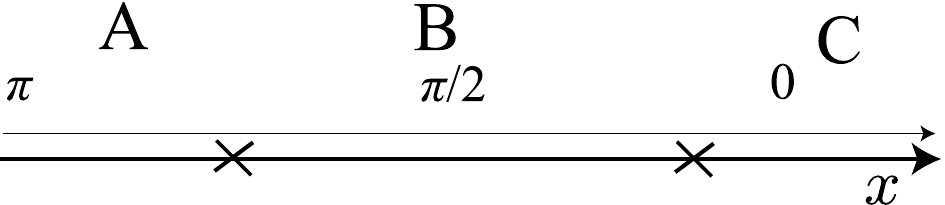}
\caption{\label{fig:analy} Spatial dependence of the phase of the pair-potential of the quasiparticle on the $X$ axis in the 
system with the half phase vortex.}
\end{figure}

\begin{figure}
\includegraphics[width = 9cm]{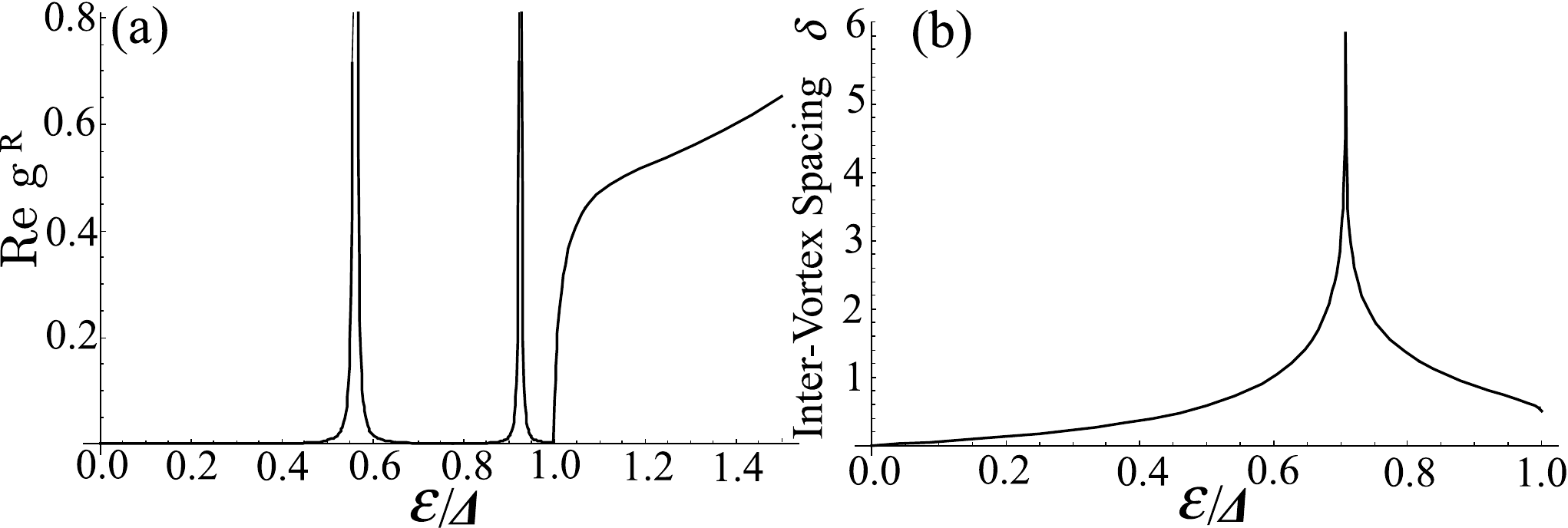}
\caption{\label{fig:fig7} Inter-vortex distance dependence of peak positions for the quasiparticle running on the $X$ axis 
in the system with the phase vortices.}
\end{figure}

We show that these novel bound states also appear in the the zero-core (phase vortex) model, 
where only the phase of the pair-potential spatially changes and the amplitude of that does not change. 
In the zero-core model, we can solve Eqs.~(\ref{eq:ar}),(\ref{eq:br}) on the $X$-axis analytically. 
We derive the $X$-line into A, B and C segments as shown in Fig.~\ref{fig:analy}. 
The initial conditions of the variables $a_{\rm I}$ and $b_{\rm I}$ of the Riccati equations (\ref{eq:ar}) and (\ref{eq:br}) are written as 
\begin{eqnarray}
a_{\rm I}(X \rightarrow - \infty) &=& e^{i \pi}(- \omega_n +\sqrt{\omega_n^2 + 1}), \\
b_{\rm I}(X \rightarrow  \infty) &=& (- \omega_n +\sqrt{\omega_n^2 + 1}).
\end{eqnarray}
Here, we use dimensionless variables as $\omega_n/\Delta_{\infty} \rightarrow \omega_n$, $X/\xi_0 \rightarrow X$.
We can solve Eqs.~(\ref{eq:ar}),(\ref{eq:br}) since the equations are $n = 2$ Bernoulli Equations in each segment. 
With the use of the particular solution in the segment B written as 
\begin{equation}
a_{\rm Bp} = e^{\frac{\pi}{2}i}(-\omega_n + \sqrt{\omega_n^2 +1}), 
\end{equation}
the general solutions in the system with the half phase vortices are written as 
\begin{eqnarray}
a_{\rm Bg}(X) &=& \frac{-i + e^{c(X+\delta)} (-1 +(1+i) \omega_n(c/2 -\omega_n))
			}{(1+e^{c(X+\delta)})c/2 +\omega_n(-1 +e^{c(X+\delta)})
                        	}, \quad  \label{eq:abg}\\
b_{\rm Bg}(X) &=& -a_{\rm Bg}(-X), \label{eq:bbg}            
\end{eqnarray}
where $c = 2 \sqrt{\omega_n^2 + 1}$.
Then, we obtain the Green function from Eqs.~(\ref{eq:greengf}),(\ref{eq:abg}) and (\ref{eq:bbg}). 
Figure \ref{fig:fig7}(a) shows the energy dependence of the retarded Green function. 
This result is consistent with the numerical calculation as shown in Fig.~\ref{fig:fig4}(a). 
The bound energy is obtained by the solution of the equation: 
\begin{equation}
i \epsilon (c' + (4 \epsilon - 4 \epsilon^{3} - c') e^{2 c' \delta}) = 0, 
\end{equation}
where  $c' = 2 \sqrt{1 - \epsilon^{2}}$ and $i \omega_{n} \rightarrow \epsilon + i \eta$ $(\eta \rightarrow +0)$. 
Therefore, the inter-vortex distance dependence of the peak position on the $X$-axis is written as
\begin{equation}
\delta = -\frac{\log (- 2 \epsilon \sqrt{1 - \epsilon^2} + 1)}{4 \sqrt{1-\epsilon^2}}.
\end{equation}
At the energy $\epsilon = 1/\sqrt{2}$, the inter-vortex distance $\delta$ diverges as shown in Fig.~\ref{fig:fig7}(b). 


\subsection{$n$ Fractional Vortices with Vorticity $1/n$}
\begin{figure}
\includegraphics[width = 6.5cm]{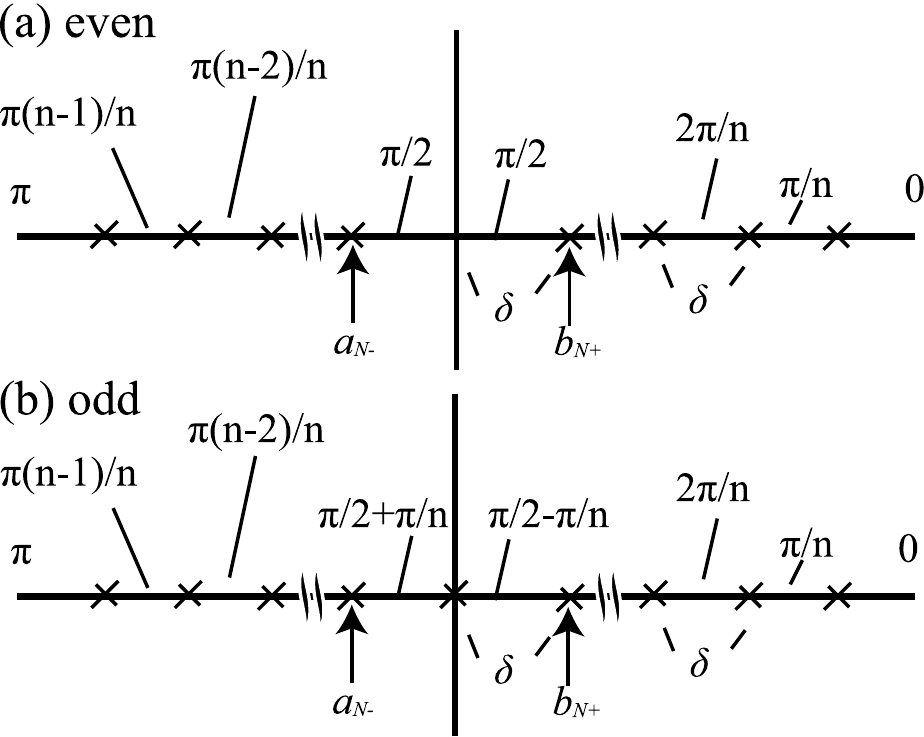}
\caption{\label{fig:fig8} Spatial dependence of the phase of the pair-potential of the quasiparticle on the $X$ axis in the 
system with the half phase vortex.}
\end{figure}

\begin{figure}
\includegraphics[width = 5cm]{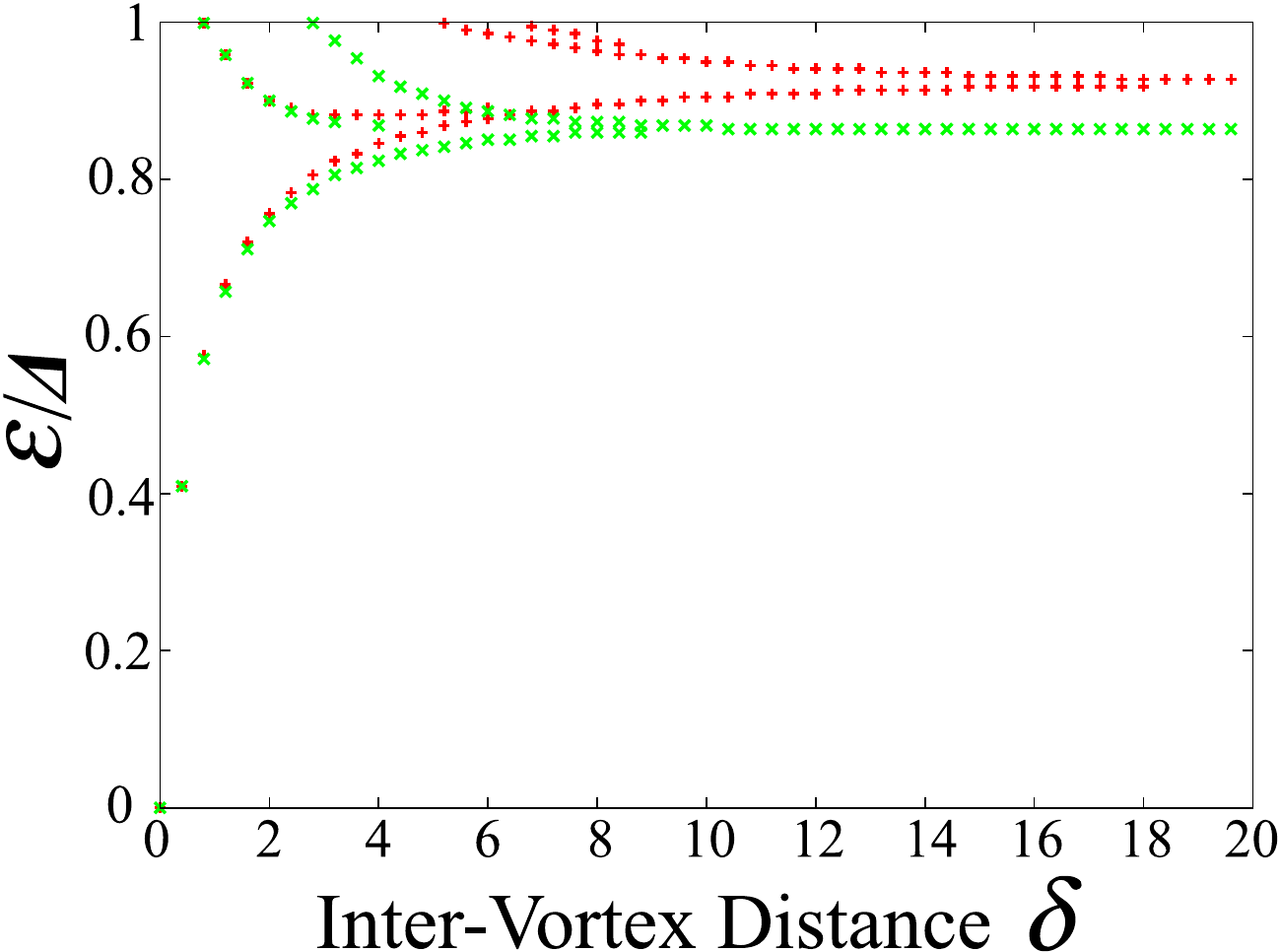}
\caption{\label{fig:fig9}(Color Online) (a)Energy dependence of the retarded Green function in the $X$-direction. (b):The inter-vortices dependence of the dispersion relation of the quasiparticle running on the $X$ axis 
in the system with the phase vortices.}
\end{figure}

We also calculate the energy dependence of the Andreev bound states in the system with $n$ fractional vortices with vorticity $1/n$ on the $X$ axis 
to investigate the effects of the vorticity.
We assume the zero-core model (phase vortex model)  
and consider the frame of the space as shown in Fig.~\ref{fig:fig8}.
Figure \ref{fig:fig9} shows the inter-vortex distance dependence of peak positions relations
 in the case of $n = 3, 4$ with the numerical calculations. 
Three and four bound states appear, respectively.
In the large inter-vortex distance limit, the bound energy is 
\begin{eqnarray}
\epsilon_{\rm 1/3,num}/\Delta &=& 0.864 \pm 0.005, \\
\epsilon_{\rm 1/4,num}/\Delta &=& 0.927 \pm 0.005.
\end{eqnarray}
As the case of the half vortices, we solve the Riccati equations (\ref{eq:ar}) and (\ref{eq:br}) analytically. 
In the long inter-vortex distance limit, the solutions $a$ and $b$ in each segment are 
equal to the bulk-solutions written as 
\begin{equation}
\partial a/\partial X = \partial b /\partial X = 0.
\end{equation}
The energy of the bound states is obtained by solving the equation written as 
\begin{eqnarray}
\sqrt{1-\epsilon^{2}} + \sqrt{1-\epsilon^{2}} \cos \frac{\pi}{n} + \epsilon \sin \frac{\pi}{n} \nonumber \\
 + i \left( - \epsilon + \sqrt{1-\epsilon^{2}} \sin \frac{\pi}{n} + \epsilon \frac{\pi}{n} \right) = 0.
\end{eqnarray}
Therefore, we obtain the energy of the bound states written as 
\begin{equation}
\epsilon = \cos \left(\frac{\pi}{2 n} \right).
\end{equation}
In the case of $n = 3,4$, the bound energies are 
\begin{eqnarray}
\epsilon_{\rm 1/3,ana} &=& \cos \left(\frac{\pi}{6} \right) \sim 0.866025 \sim \epsilon_{\rm 1/3,num}, \\
\epsilon_{\rm 1/4,ana} &=& \cos \left(\frac{\pi}{8} \right) \sim 0.92388 \sim \epsilon_{\rm 1/4,num},
\end{eqnarray}
respectively.

These results can be understood by the analogy of the Josephson junctions.
In this limit, 
the quasiparticles running on $X$ axis pass through the single vortex with vorticity $1/n$.
Therefore, we can regard this single vortex as 
 the single Josephson junction between the superconductors with different phase of the pair-potential. 
The bound energy at the junction between two superconductors with the phase difference $\Delta \phi = \pi/n$ 
is written as 
\begin{equation}
\epsilon = \cos \left( \frac{\Delta \phi}{2} \right) = \cos \left( \frac{\pi}{2 n } \right).
\end{equation}
\section{Connected Fractional Vortices}
\subsection{Two Half Vortices}
We consider the pair function of the two connected half fractional vortices is written as 
\begin{eqnarray}
\Delta(r) &=& f_{0}(X,Y)  \exp[i (\phi_+(X,Y)+\phi_-(X,Y))], \nonumber \\ \\
	&\equiv& f_{0}(r) \exp [i \phi(r)], 
\end{eqnarray}
where 
\begin{eqnarray}
f_{0}(X,Y) &=& \left\{
\begin{array}{ll}
\sqrt{ \tanh (|Y|) f_{+}(X,Y) }, &\quad - X_+ < X <0 \\
\sqrt{f_{-}(X,Y)  \tanh (|Y|) },  &\quad  0 < X < X_+\\
\sqrt{ f_{-}(X,Y) f_{+}(X,Y) },  &\quad |X| > X_+ 
\end{array}
\right.\nonumber \\
f_{\pm}(X,Y) &=& \tanh (\sqrt{Y^2 + (X \mp X_+)^2})\\
\phi_{\pm}(X,Y) &=& \arctan(Y/(X \pm \delta))/2. 
\end{eqnarray}
As shown in Fig.~\ref{fig:gap}(b), the gap amplitude on the branch cut is zero. 
\subsubsection{Local Density of States}
\begin{figure}
\includegraphics[width = 7cm]{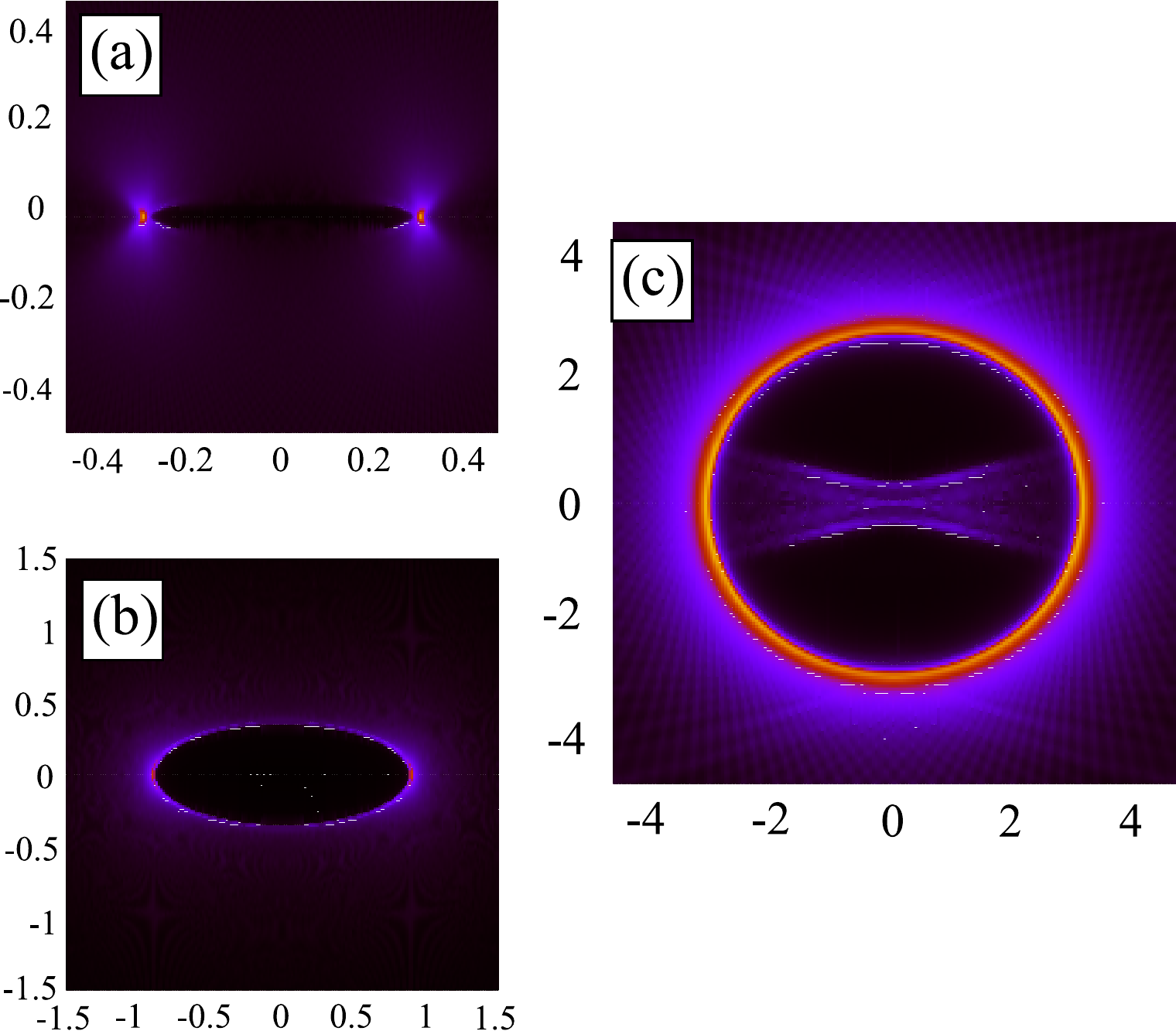}
\caption{\label{fig:conepfv}(Color Online) Local density of the states in the system with two connected half 
fractional vortices. 
The inter-vortex distance $\delta = 0.8 \xi_0$. 
(a):the energy $\epsilon = 0.1 \Delta_{\infty}$, (b): the energy $\epsilon = 0.4 \Delta_{\infty}$
 and (c):the energy $\epsilon = 0.9 \Delta_{\infty}$. the smearing factor $\eta = 0.001\Delta_{\infty}$.}
\end{figure}

We calculate the local density of states with two connected half vortices. 
As shown in Fig.~\ref{fig:conepfv}, the LDOS with two connected half vortices is similar to that with two isolated half vortices shown in Fig.~\ref{fig:fig2} qualitatively. 
This result suggests that the appearance of the bow-tie-shaped bound states in the high energy region does not depend on the details of the distribution of the 
gap amplitude. 
\subsubsection{Energy dependence of the bound states}
\begin{figure}
\includegraphics[width = 8cm]{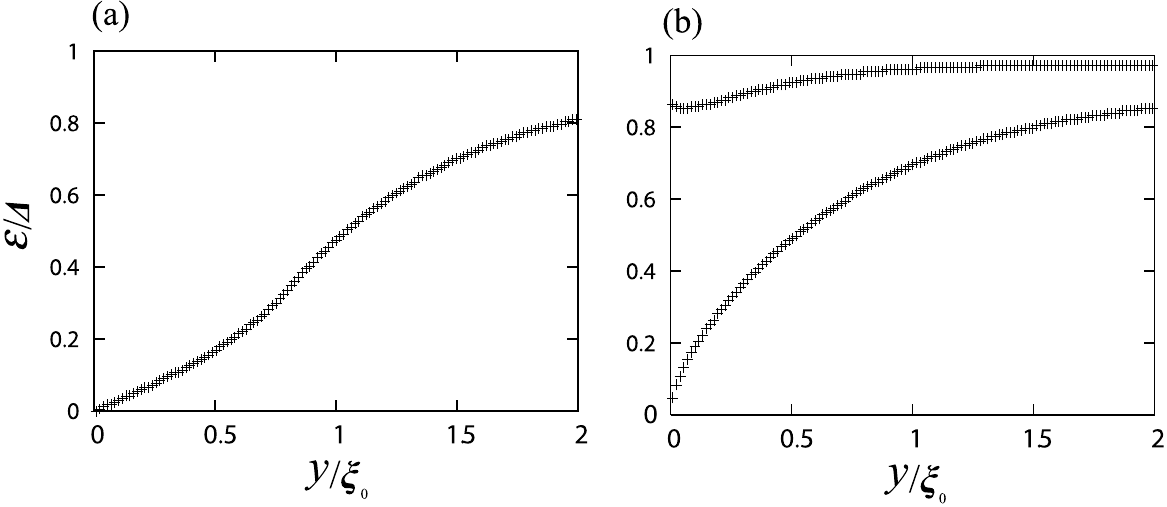}
\caption{\label{fig:dis}Energy dependence of peak positions for the quasiparticle running in the (a):$Y$ direction and (b):$X$ direction with connected two half vortices.}
\end{figure}

We obtain the energy dependence of the bound states as shown in Fig.~\ref{fig:dis}. 
\begin{figure}
\includegraphics[width = 8cm]{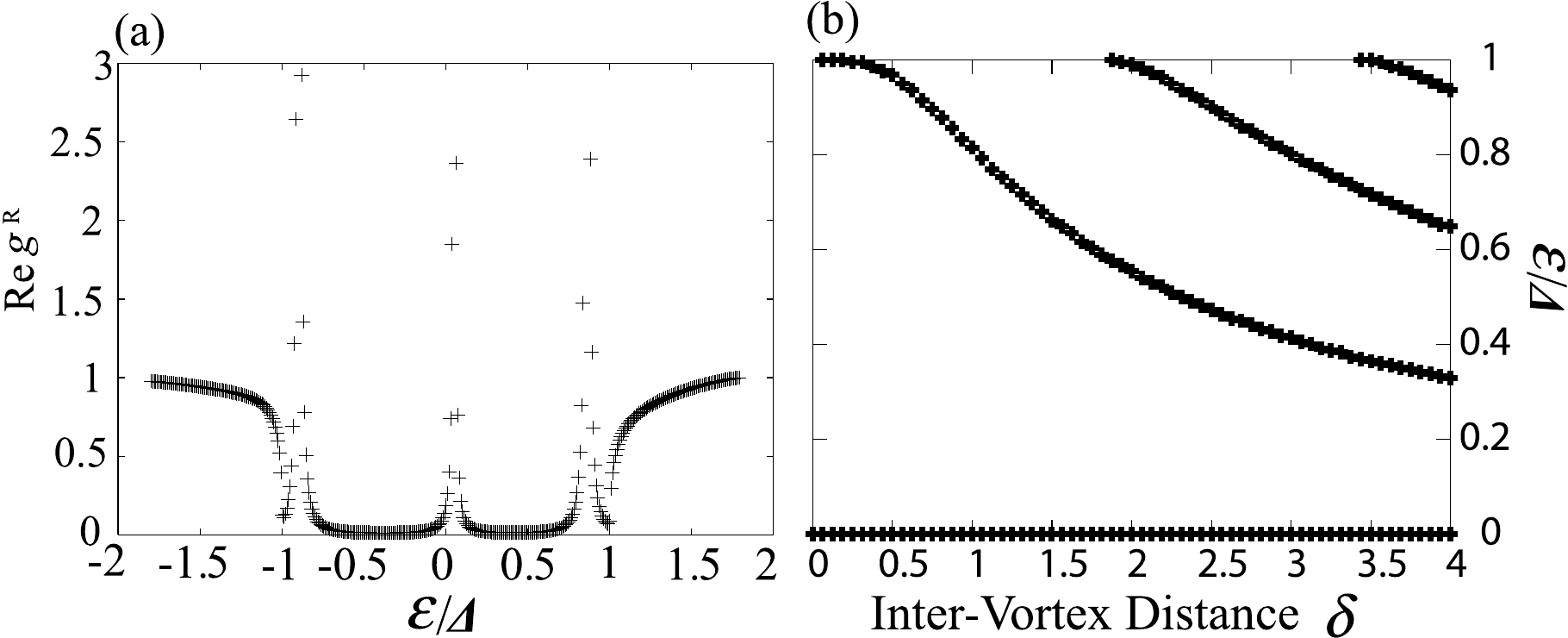}
\caption{\label{fig:condel}(a):Energy dependence of the retarded Green function in the $X$-direction near the origin ($y = 0.01 \xi_{0}$). 
(b) the inter-vortex distance dependence of the energy of the bound states.}
\end{figure}
The energy dependence of peak positions for the quasiparticles running in the $Y$-direction with two connected half vortices is similar to 
that with two isolated half vortices. 
On the other hand, 
the quasiparticles running in the $X$-direction do not have the gap. 
The bound states for these quasiparticles appear in the high-energy region as shown in Fig.~\ref{fig:condel}(a).  
The pair function on the $Y$ axis with connected vortices is similar to that with isolated vortices. 
On the $X$ axis, however, the pair function with connected vortices is zero on the segment between two vortices. 
The inter-vortex distance dependence of the bound energy for these quasiparticles is shown in Fig.~\ref{fig:condel}(b). 
With increasing the inter-vortex distance $\delta$, the several bound states appear. 
For the quasiparticles running on $X$-axis, such a pair amplitude is similar to that of the S/N/S  junction where the length of the normal state $L$ 
is $L = 2 \delta$. 
Therefore, many bound states appear when the length of the normal state region $L$ becomes large.

%
\section{Discussion}

We do not solve the gap equation self-consistently for simplicity. 
In the real materials, we do not know whether the pair-amplitude on the branch cut is zero (isolated vortices) or 
not (connected vortices).  
However, we show that the appearance of the bow-tie-shaped LDOS pattern in the high energy region 
is robust against the spatial dependence of the pair-amplitude. 
This result suggests that the topological effects of the fractional vortices cause the bow-tie-shaped LDOS pattern 
in the high energy region.

With the use of quasiclassical Eilenberger theory, we can 
separately consider each quasiparticles with certain momentum $k$. 
Therefore, we can suggest that the quasiparticles running through both half vortices 
form the novel bound states in the high energy region ($\Delta_{\infty} \gtrsim \epsilon \gtrsim 0.8 \Delta_{\infty}$) in unconventional superconductors.

\section{Conclusion}
In conclusion, we investigated the local density of states around two half vortices.
We studied the isolated and connected two half vortices, respectively. 
We obtained the novel bound states in the high energy region in both cases. 
With increasing energy, the LDOS pattern changes from the triangular-like pattern to the ellipsoidal pattern.
In the high energy region, the novel bow-tie-shaped LDOS pattern is inside the circular LDOS pattern. 
We showed that the LDOS patterns by the numerical method can be understood by the enveloping curve of the quasiparticle paths.
We also showed that the appearance of these novel bound states does not depend on the details of the gap-amplitude distribution. 
These results suggest that these novel LDOS pattern can be observed by the STM/STS. 
We also calculate the quasiparticle dispersion in the system with isolated $n$ fractional vortices with $1/n$ vorticity. 
We noted that $n$ fractional vortices with $1/n$ vorticity is similar to the multi Josephson junctions.

\section*{Acknowledgment}
We thank Y. Higashi, N. Hayashi, M. Ichioka and K. Machida for helpful discussions.
The calculations were partially performed by the supercomputing systems SGI ICE X at the Japan Atomic Energy Agency. Y. N. was partially supported by 
JSPS KAKENHI Grant Number 18K11345 and 18K03552, the ``Topological Materials Science'' (No. JP18H04228) KAKENHI on Innovative Areas from JSPS of Japan. 


\end{document}